\documentclass[aps,prd,%multicol,
twocolumn,
preprintnumbers,groupedaddress]{revtex4}
\usepackage{epsbox}
\usepackage{amsmath}
\usepackage[dvips]{graphicx}

\begin{document}
\preprint{\vtop{{\hbox{YITP-06-15}\vskip-0pt
                 \hbox{KANAZAWA-06-04} \vskip-0pt
%                 \hbox{hep-ph/0604207} 
}}}
%\date{\today}

\vspace{20mm}

\title{
Production of neutral and doubly charged partners 
of ${\bf D_{s0}^+(2317)}$ 
}
\author{Kunihiko Terasaki}
%\author{}
%\email[e-mail: ]{terasaki@hep.s.kanazawa-u.ac.jp,
%terasaki@yukawa.kyoto-u.ac.jp}
\affiliation{Yukawa Institute for Theoretical Physics, Kyoto University,
Kyoto 606-8502, Japan\\
Institute for Theoretical Physics, Kanazawa University, 
Kanazawa 920-1192, Japan}
%\numberwithin{equation}{section}
\thispagestyle{empty}

\begin{abstract}
Production of the neutral and doubly charged partners $\hat F_I^0$ and 
$\hat F_I^{++}$ of $ D_{s0}^+(2317)$ as the charm-strange four-quark 
meson $\hat F_I^{+}$ is studied in relation to observation of 
$D_{s0}^+(2317)$. It is argued that observation of 
$\hat F_I^{++},\,\hat F_I^{0}$ and $\hat F_0^{+}$ in inclusive 
$e^+e^-\rightarrow c\bar c$ would be difficult, although 
$\hat F_I^{++}$ might be observed in $B_u^+\rightarrow D_s^+\pi^+D^-$. 
Observations of $\hat F_I^0$ and $\hat F_0^+$ might be possible in 
hadronic $B$ decays. 
\end{abstract}
\maketitle

Recently, the resonance $D_{s0}^+(2317)$, which decays into 
$D_s^+\pi^0$, has been observed in $e^+e^-\rightarrow c\bar c$ 
experiments~\cite{BABAR-D_{s0},CLEO-D_{s0}}. In addition, a resonance 
that is degenerate with it has been observed in $B$ 
decays~\cite{BELLE-D_{s0},BABAR-D_{s0}-B}. 
While it is known that their spin-parity is $J^P = 0^+$ and their 
width is very narrow, their isospin quantum number has not yet been 
definitively determined. 
(A comprehensive review on new heavy mesons is given in 
Ref.~\cite{Swanson}.) 
To determine the isospin quantum number of 
$D_{s0}^+(2317)$, its decay properties have been 
studied~\cite{HT-isospin} by assigning it to various scalar meson 
states: 
(i) the $I_3=0$ component $\hat F_I^+$ of iso-triplet four-quark 
mesons~\cite{Terasaki-D_{s0}}, 
%%%%%%%%%%%%%%%%%%%%%%%%%%%%%%%%%%%%%%%%%%%%%%%%%%%%%%%%%%%%%%%%%%%%%%%%
$\hat F_I \sim [cn][\bar s\bar n]_{I=1}\,
(n=u,\,d)$; 
%%%%%%%%%%%%%%%%%%%%%%%%%%%%%%%%%%%%%%%%%%%%%%%%%%%%%%%%%%%%%%%%%%%%%%%%
(ii) the iso-singlet four-quark meson,
%%%%%%%%%%%%%%%%%%%%%%%%%%%%%%%%%%%%%%%%%%%%%%%%%%%%%%%%%%%%%%%%%%%%%%%%
$\hat F_0 \sim [cn][\bar s\bar n]_{I=0}$,
%%%%%%%%%%%%%%%%%%%%%%%%%%%%%%%%%%%%%%%%%%%%%%%%%%%%%%%%%%%%%%%%%%%%%%%%
which might not be identical to that considered in Ref.~\cite{CH};  
(iii) the conventional scalar $D_{s0}^{*+}\sim \{c\bar s\}$ 
meson~\cite{DGG}. 
The results obtained in these studies are 
%%%%%%%%%%%%%%%%%%%%%%%%%%%%%%%%%%%%%%%%%%%%%%%%%%%%%%%%%%%%%%%%%%%%%%%%%%
(i) $R(\hat F_I^+)\simeq 0.005$, (ii) $R(\hat F_0^+)\simeq 7$ and  
(iii) $R(\hat D_{s0}^{*+})\simeq 60$,  
%%%%%%%%%%%%%%%%%%%%%%%%%%%%%%%%%%%%%%%%%%%%%%%%%%%%%%%%%%%%%%%%%%%%%%%%%%
where $R(S)$ is given by 
%%%%%%%%%%%%%%%%%%%%%%%%%%%%%%%%%%%%%%%%%%%%%%%%%%%%%%%%%%%%%%%%%%%
$R(S)= {\Gamma(S \rightarrow D_{s}^{*+}\gamma)}/
       {\Gamma(S \rightarrow D_{s}^{+}\pi^0)}$, 
%%%%%%%%%%%%%%%%%%%%%%%%%%%%%%%%%%%%%%%%%%%%%%%%%%%%%%%%%%%%%%%%%%%
with $S=\hat F_I^+,\,\hat F_0^+,\,D_{s0}^{*+}$. The same approach 
predicts~\cite{HT-isospin} $R(D_{s}^{*+})^{-1}\simeq 0.06$ which 
reproduces well the measured ratio~\cite{BABAR-Radiative},    
%%%%%%%%%%%%%%%%%%%%%%%%%%%%%%%%%%%%%%%%%%%%%%%%%%%%%%%%%%%%%%%%%%%
$R(D_{s}^{*+})^{-1}_{\rm exp} = 0.062\pm 0.006\pm 0.005$, 
%%%%%%%%%%%%%%%%%%%%%%%%%%%%%%%%%%%%%%%%%%%%%%%%%%%%%%%%%%%%%%%%%%%
where $S=D_s^{*+}$, and thus the present approach seems to be 
sufficiently reliable. By comparing the above results with the 
experimental constraint~\cite{CLEO-D_{s0}} 
%%%%%%%%%%%%%%%%%%%%%%%%%%%%%%%%%%%%%%%%%%%%%%%%%%%%%%%%%%%%%%%%%%%
\begin{equation}
R(D_{s0}^+(2317))_{\rm exp} 
 < 0.059, 
                                                  \label{eq:CLEO}
\end{equation}
%%%%%%%%%%%%%%%%%%%%%%%%%%%%%%%%%%%%%%%%%%%%%%%%%%%%%%%%%%%%%%%%%%%
it has been concluded that experiments favor the assignment (i) 
over (ii) and (iii). Note that its assignment to an iso-singlet 
$\{DK\}$ molecule,~\cite{BCL} as an additional possibility, has 
already been rejected,~\cite{MS} because it leads to the relation 
$R(\{DK\}) \gg R(D_{s0}^+(2317))_{\rm exp}$.  

From the above considerations, we see that it is natural to assign 
$D_{s0}^+(2317)$ to $\hat F_I^+$. However, its neutral and doubly 
charged partners, $\hat F_I^0$ and $\hat F_I^{++}$, have not yet 
been observed 
experimentally~\cite{BABAR-search, CDF-search, CLEO-search}. 
With this in mind, in this short note, we study the production of 
charm-strange scalar four-quark mesons ($\hat F_I^{++,+,0}$ and 
$\hat F_0^+$) in relation to the observation of $D_{s0}^+(2317)$ by 
assigning it to $\hat F_I^+$ and discuss why experiments have 
observed $D_{s0}^+(2317)$ but not its neutral and doubly charged 
partners. To this end, we consider their production 
through weak interactions as a possible mechanism, because OZI-rule 
violating productions of multiple $q\bar q$ pairs and their 
recombinations into four-quark meson states are believed to be 
strongly suppressed at high energies. (Such multiple $q\bar q$ pair 
creations might produce backgrounds of four-quark meson signals.) 
First, we construct quark-line diagrams within the minimal (i.e., one) 
$q\bar q$ pair creation, noting the OZI rule. 
%%%%%%%%%%%%%%%%%%%%%%%%%%%%%%%%%%%%%%%%%%%%%%%%%%%%%%%%%%%%%%%%%%%%%%%%%%
\begin{figure}[!b]  %
%\begin{center}\hspace{-0mm}
\includegraphics[width=75mm,clip]{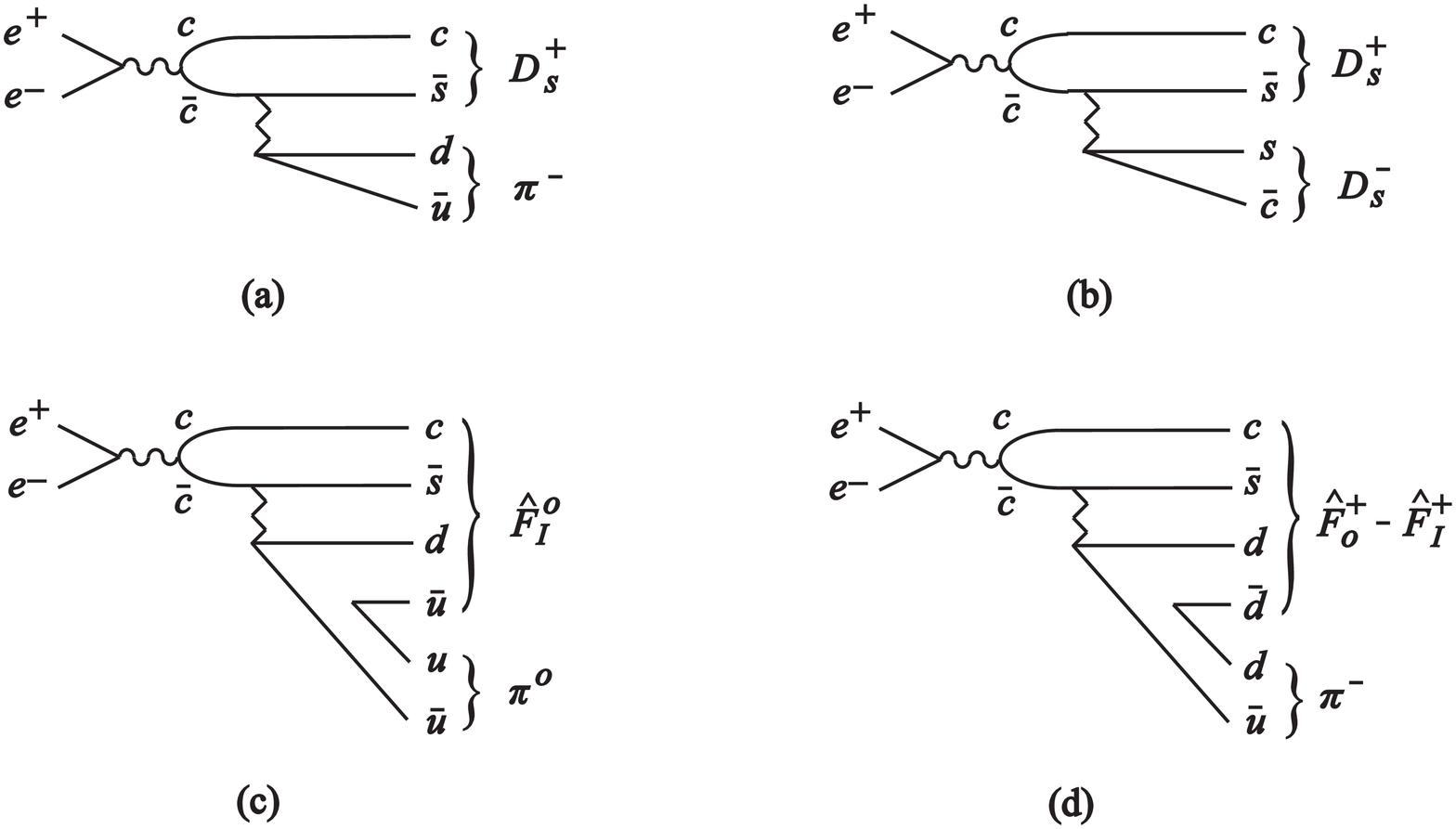}
\label{fig:c-cbar-8.eps}       %
%\end{center} 
\begin{quote}  
%\caption{
Fig.~1. Production of charm-strange scalar mesons through 
$e^+e^-\rightarrow c\bar c$ within the minimal quark-antiquark pair 
creation. (a) and (b) describe the production of $D_s^+\pi^-$, 
$D_s^{*+}\pi^-$, $D_s^+\rho^-$, etc. and $D_s^+D_s^-$, $D_s^{*+}D_s^-$, 
$D_s^+D_s^{*-}$, etc., respectively. The production of $\hat F_I^0\pi^0$ 
and $(\hat F_0^+,\,\hat F_I^+)\pi^-$ is described by (c) and (d), 
respectively.  
%}
\end{quote}
\end{figure}%    
%%%%%%%%%%%%%%%%%%%%%%%%%%%%%%%%%%%%%%%%%%%%%%%%%%%%%%%%%%%%%%%%%%%%%%
Because there is no diagram yielding $\hat F_I^{++}$ production in 
this approximation, as seen in Fig.~1, it is easy to understand why  
the BABAR and CLEO experiments did not find any evidence of 
$\hat F_I^{++}$ in $e^+e^-\rightarrow c\bar c$. Production of 
iso-triplet $\hat F_I^{+,0}$ and iso-singlet $\hat F_0^+$ mesons 
through $e^+e^-\rightarrow c\bar c$ results from the processes whose 
diagrams are displayed in Figs.~1(c) and (d). The diagrams Figs.~1(a) 
and (b) describe productions of $D_s^+\pi^-$, $D_s^{*+}\pi^-$, 
$D_s^+\rho^-$, etc., and $D_s^+D_s^-$, $D_s^+D_s^{*-}$, 
$D_s^{*+}D_s^-$, etc. Their weak vertices are given by the color 
favored spectator diagrams. It is known that such a spectator decay, 
whose amplitude is proportional to $a_1$, is much stronger than a 
color mismatched decay, whose amplitude is proportional to $a_2$ 
(explicitly, we have $|a_2/a_1|^2 \simeq 6.8\times 10^{-3}$ at the 
scale of the charm mass~\cite{Neubert}), as long as non-factorizable 
contributions are ignored. Here, $a_1$ and $a_2$ are the coefficients 
of the four-quark operators given by products of charged currents and 
neutral currents, respectively, in the effective weak Hamiltonian 
after a Fierz reshuffling. In hadronic weak decays of $B$ mesons, 
non-factorizable contributions are actually small, and they are much 
smaller at higher energies. Therefore, very large numbers of 
$D_s^+\pi^-$ events, 
which are produced through a reaction described by Fig.~1(a) (and 
semi-inclusive $e^+e^-\rightarrow c\bar c \rightarrow D_s^+\pi^- + X$), 
would obscure the signal of $\hat F_I^0 \rightarrow D_s^+\pi^-$ events,  
which are produced through Fig.~1(c). 
The latter involves rearrangements of colors,  
as in color mismatched decays, and is much more strongly suppressed 
than the color favored ones, as we see in the case of $B$ decays below. 
For this reason, it is not easy to extract the 
$\hat F_I^0 \rightarrow D_s^+\pi^-$ signals in inclusive 
$e^+e^-\rightarrow c\bar c$ experiments. Noting these points, 
it is understood why the CLEO~\cite{CLEO-search} and 
BABAR~\cite{BABAR-search} experiments found no signal of 
$\hat F_I^0$ and  $\hat F_I^{++}$. In the case of $\hat F_I^+$, 
however, there do not 
exist large numbers of background events described by Figs.~1(a) and 
(b) because its main decay is $\hat F_I^+\rightarrow D_s^+\pi^0$. 
In fact, BABAR~\cite{BABAR-D_{s0}} and CLEO~\cite{CLEO-D_{s0}} have 
observed $D_{s0}^+(2317)\rightarrow D_s^+\pi^0$. This seems to imply 
that the production of four-quark mesons in hadronic weak decays plays 
an essential role. Figures 1(c) and (d) describe the creation of 
$\hat F_I^0\pi^0$ and $\hat F_{I,0}^+\pi^-$, respectively. 
The iso-triplet $\hat F_I^+$ decays dominantly into $D_s^+\pi^0$, but 
its decay into $D_s^{*+}\gamma$ is strongly suppressed, as discussed 
above. Therefore, it is easily understood experiments have observed 
$D_{s0}^+(2317)$ in the $D_s^+\pi^0$ channel but not in the 
$D_s^{*+}\gamma$ channel. Figure 1(d) includes the production of 
$\hat F_0^+$, which can decay much more strongly into $D_s^{*+}\gamma$ 
than $D_s^+\pi^0$, as mentioned above. 
Therefore, it is natural to conjecture that reconstruction of 
$\hat F_0^+\rightarrow D_s^{*+}\gamma$ might be more efficient 
as a method to search for $\hat F_0^+$. 
%%%%%%%%%%%%%%%%%%%%%%%%%%%%%%%%%%%%%%%%%%%%%%%%%%%%%%%%%%%%%%%%%%%%%%%%%%
\begin{figure}[!b]     %
\begin{center}  %\hspace{-20mm}
\includegraphics[width=70mm,clip]{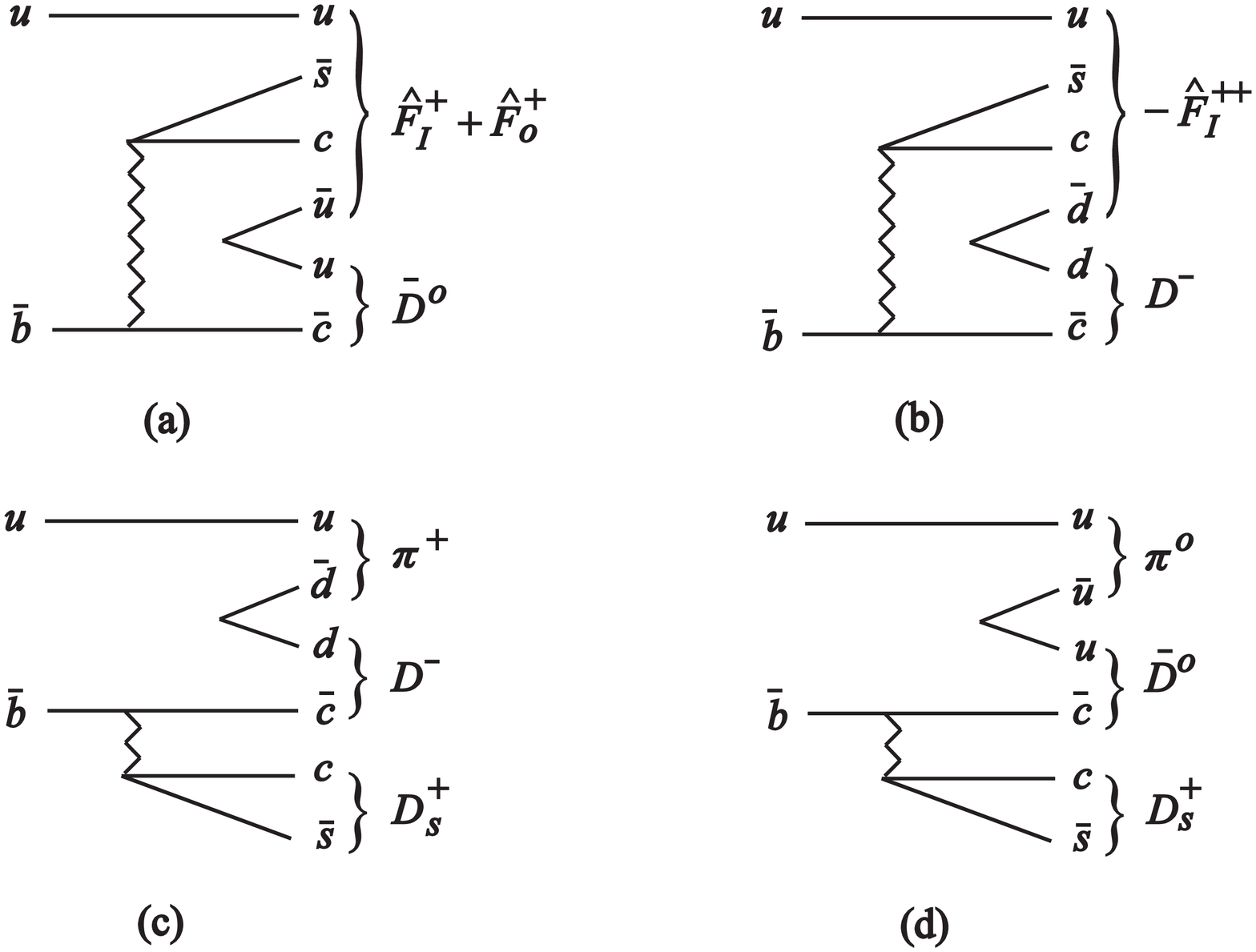}
\label{fig:Bu-8.eps}              %
\end{center} 
\begin{quote}  
%\caption{
Fig.~2. Production of charm-strange scalar mesons in weak decays of the 
$B_u$ meson. (a) describes the production of $\hat F_I^+$ and 
$\hat F_0^+$ with $\bar D^0\,({\rm or}\,\bar D^{*0})$, 
(b) the production of $\hat F_I^{++}$ with $D^-\,({\rm or}\,D^{*-})$, and  
(c) and (d) the production of $D_{s}^+\pi^+$ with $D^-$ and 
$D_{s}^+\pi^0$ with $\bar D^0$, respectively.
%}
\end{quote}
\end{figure}%          
%%%%%%%%%%%%%%%%%%%%%%%%%%%%%%%%%%%%%%%%%%%%%%%%%%%%%%%%%%%%%%%%%%%%%%%
However, $D_s^{*+}$ and $\gamma$ (from $D_s^{*-}\rightarrow D_s^-\gamma$) 
produced in the spectator diagrams Figs.~1(a) and (b) (and in 
$e^+e^-\rightarrow c\bar c\rightarrow D_s^{*+}D_s^-$, etc., through 
strong interactions) obscure the above signal of $D_s^{*+}\gamma$. 
Hence it is clear why experiments have observed no scalar resonance in 
the $D_s^{*+}\gamma$ channel. 

To search for $\hat F_I^0$ in $e^+e^-\rightarrow c\bar c$ experiments, 
it might be necessary to study an exclusive 
%%%%%%%%%%%%%%%%%%%%%%%%%%%%%%%%%%%%%%%%%%%%%%%%%%%%%%%%%%%%%%%%%%%%%%%%
$e^+e^-\rightarrow c\bar c\rightarrow D_s^+\pi^-\pi^0$ 
%%%%%%%%%%%%%%%%%%%%%%%%%%%%%%%%%%%%%%%%%%%%%%%%%%%%%%%%%%%%%%%%%%%%%%%%
reaction, depicted in Fig.~1(c).   
Similarly to the situation discussed above, it might be possible to 
observe  $\hat F_0^+$ by analyzing an exclusive 
%%%%%%%%%%%%%%%%%%%%%%%%%%%%%%%%%%%%%%%%%%%%%%%%%%%%%%%%%%%%%%%%%%%%%%%
$e^+e^- \rightarrow D_{s}^{*+}\pi^-\gamma 
\rightarrow D_{s}^+\pi^-\gamma\gamma$ 
%%%%%%%%%%%%%%%%%%%%%%%%%%%%%%%%%%%%%%%%%%%%%%%%%%%%%%%%%%%%%%%%%%%%%%%
reaction. To get rid of the large numbers of background events from 
many other channels, it seems that analyses of exclusive reactions 
are important; i.e., it would be difficult to pick out the signals of 
$\hat F_I^0\rightarrow D_s^+\pi^-$ and 
$\hat F_0^+\rightarrow D_s^{*+}\gamma$ events 
%%%%%%%%%%%%%%%%%%%%%%%%%%%%%%%%%%%%%%%%%%%%%%%%%%%%%%%%%%%%%%%%%%%%%%
if $D_s^+\pi^-$ and $D_s^+\gamma\gamma$ events were collected 
inclusively. 
%%%%%%%%%%%%%%%%%%%%%%%%%%%%%%%%%%%%%%%%%%%%%%%%%%%%%%%%%%%%%%%%%%%%%%%%%%
\begin{figure}[!b]    %
%\begin{center}  %\hspace{-20mm}
\includegraphics[width=70mm,clip]{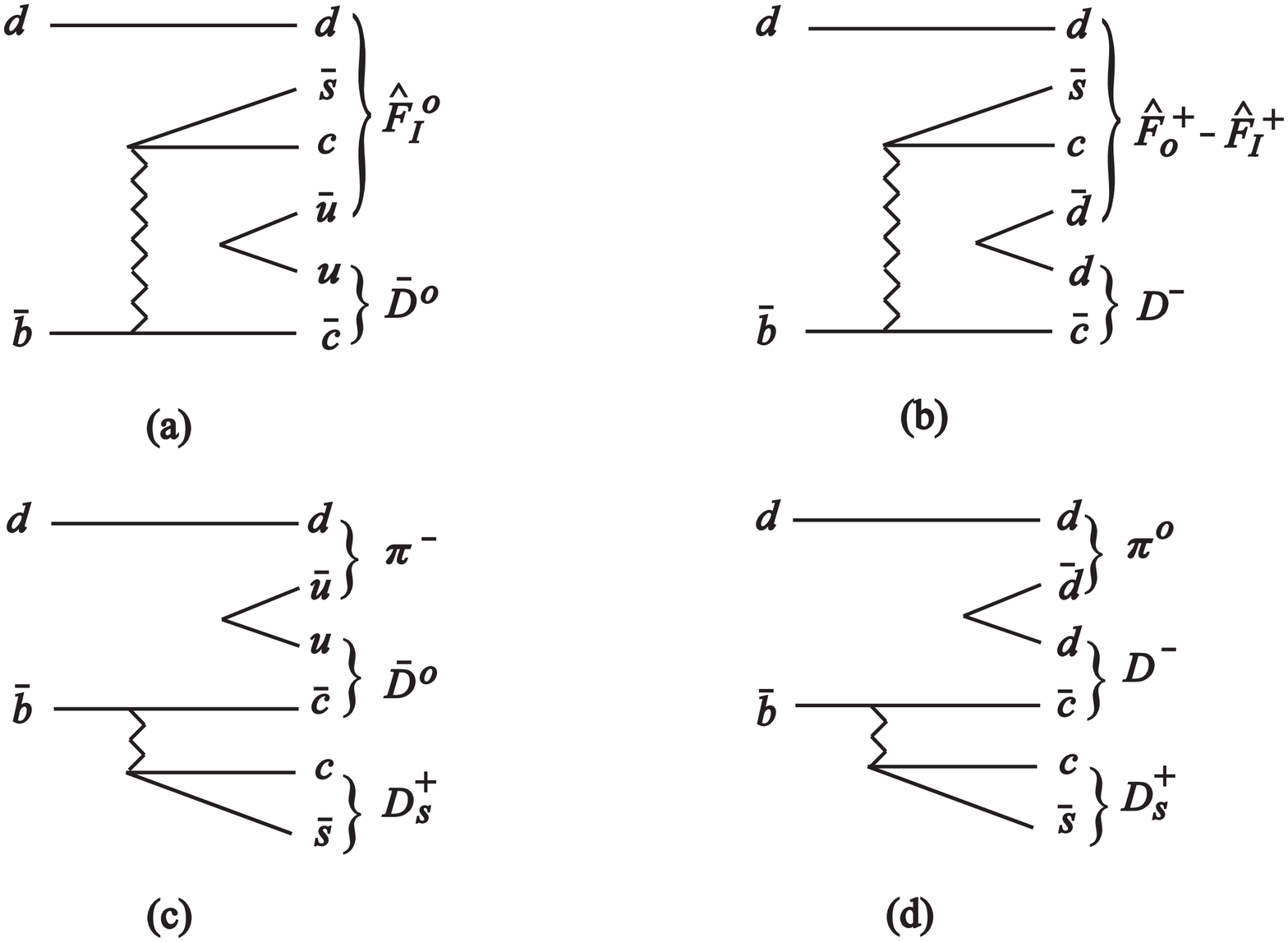}
\label{fig:Bd-8.eps}             %
%\end{center} 
%\caption{      %
\begin{quote} 
Fig.~3. Production of charm-strange scalar mesons in weak decays of the 
$B_d$ meson. (a) describes the production of $\hat F_I^0$ with $\bar D^0$ 
(or $\bar D^{*0}$), (b) the production of $\hat F_I^+$ and $\hat F_0^+$ 
with $D^-$ (or $D^{*-}$). (c) and (d) depict the production of 
$D_s^+\pi^-$ with $\bar D^0$ and $D_s^+\pi^0$ with $D^-$, respectively.    
%} 
\end{quote}    
\end{figure}%        
%%%%%%%%%%%%%%%%%%%%%%%%%%%%%%%%%%%%%%%%%%%%%%%%%%%%%%%%%%%%%%%%%%%%%%%%%

Because it is difficult to observe $\hat F_I^{++}$ in 
$e^+e^-\rightarrow c\bar c$ experiments, as discussed above, we now 
study the production of charm-strange scalar four-quark mesons, 
$\hat F_I^{++,+,0}$ and $\hat F_0^{+}$, in $B$ decays.  
For this purpose, we again draw 
quark-line diagrams describing such production within the minimal 
$q\bar q$ pair creation, as above. As expected from Figs.~2 and 3, 
a resonance peak that is approximately degenerate with $D_{s0}^+(2317)$ 
has been observed in $B$ decays: 
%%%%%%%%%%%%%%%%%%%%%%%%%%%%%%%%%%%%%%%%%%%%%%%%%%%%%%%%%%%%%%%%%%%%%%%%%
$B_u^+\rightarrow \bar D^0\tilde D_{s0}^+(2317)
[D_s^+\pi^0, D_s^{*+}\gamma]$ 
and 
$B_d^0\rightarrow D^-\tilde D_{s0}^+(2317)[D_s^+\pi^0, D_s^{*+}\gamma]$  
%%%%%%%%%%%%%%%%%%%%%%%%%%%%%%%%%%%%%%%%%%%%%%%%%%%%%%%%%%%%%%%%%%%%%%%%%
in the BELLE experiment~\cite{BELLE-D_{s0}}, and 
%%%%%%%%%%%%%%%%%%%%%%%%%%%%%%%%%%%%%%%%%%%%%%%%%%%%%%%%%%%%%%%%%%%%%%%%%
$B_u^+\rightarrow 
\bar D^0({\rm or}\,\bar D^{*0})\tilde D_{s0}^+(2317)[D_s^+\pi^0]$, 
and $B_d^0\rightarrow 
D^-({\rm or}\,D^{*-})\tilde D_{s0}^+(2317)[D_s^+\pi^0]$ 
%%%%%%%%%%%%%%%%%%%%%%%%%%%%%%%%%%%%%%%%%%%%%%%%%%%%%%%%%%%%%%%%%%%%%%%%%
in the BABAR experiment~\cite{BABAR-D_{s0}-B}. Here, the new resonance 
has been denoted  by $\tilde D_{s0}^+(2317)$ to distinguish it from the 
previous $D_{s0}^+(2317)$, although it is usually identified with 
$D_{s0}^+(2317)$. This is because the BELLE collaboration observed 
signals that may correspond to the new resonance in both the $D_s^+\pi^0$ 
and $D_s^{*+}\gamma$ channels. This is quite different from the case in 
$e^+e^-\rightarrow c\bar c$ experiments, and therefore it might not be 
identical to $D_{s0}^+(2317)$, although their masses are approximately 
equal. The decays mentioned above can proceed through Figs.~2(a) and 3(b), 
and hence the new resonance can be assigned to $\hat F_I^+$ when it is 
observed in the $D_s^+\pi^0$ channel, while it might be assigned to 
$\hat F_0^+$ when it is observed in the $D_s^{*+}\gamma$ channel, because 
these diagrams involve both $\hat F_I^+$ and $\hat F_0^+$, whose main 
decays are quite different from each other. 

As $\tilde D_{s0}^+(2317)$ has been observed in the 
%%%%%%%%%%%%%%%%%%%%%%%%%%%%%%%%%%%%%%%%%%%%%%%%%%%%%%%%%%%%%%%%%%%%%
$B_u^+\rightarrow 
\bar D^0({\rm or}\,\bar D^{*0})\tilde D_{s0}^+(2317)[D_s^+\pi^0]$ 
%%%%%%%%%%%%%%%%%%%%%%%%%%%%%%%%%%%%%%%%%%%%%%%%%%%%%%%%%%%%%%%%%%%%%%%%
decay, which is depicted by the diagram in Fig. 2(a), observations of 
$\hat F_I^{++}$ and $\hat F_I^0$ are expected in the process 
%%%%%%%%%%%%%%%%%%%%%%%%%%%%%%%%%%%%%%%%%%%%%%%%%%%%%%%%%%%%%%%%%%%%%%%%
$B_u^+\rightarrow D^-({\rm or}\,D^{*-})\hat F_I^{++}[D_s^+\pi^+]$, 
%%%%%%%%%%%%%%%%%%%%%%%%%%%%%%%%%%%%%%%%%%%%%%%%%%%%%%%%%%%%%%%%%%%%%%%%
as shown in Fig.~2(b), and in the process 
%%%%%%%%%%%%%%%%%%%%%%%%%%%%%%%%%%%%%%%%%%%%%%%%%%%%%%%%%%%%%%%%%%%%%%%%
$B_d^0\rightarrow \bar D^0\hat F_I^0[D_s^+\pi^-]$, 
%%%%%%%%%%%%%%%%%%%%%%%%%%%%%%%%%%%%%%%%%%%%%%%%%%%%%%%%%%%%%%%%%%%%%%%%
as shown in Fig.~3(a), respectively. However, amplitudes for the 
production of $\hat F_I^0$ depicted in Figs.~4(a) and (b) interfere 
destructively, because the anti-symmetry property of the 
$[cd][\bar u\bar s]$ wavefunction leads to opposite signs for the 
$\hat F_I^0$ phases in these diagrams. In addition, the spectator decays 
described by Fig.~4(d) lead to productions of large numbers of 
background $D_s^+\pi^-$ events. 
%%%%%%%%%%%%%%%%%%%%%%%%%%%%%%%%%%%%%%%%%%%%%%%%%%%%%%%%%%%%%%%%%%%%%%%%
\begin{figure}[!b]    %
%\begin{center}     %\hspace{-20mm}
\includegraphics[width=70mm,clip]{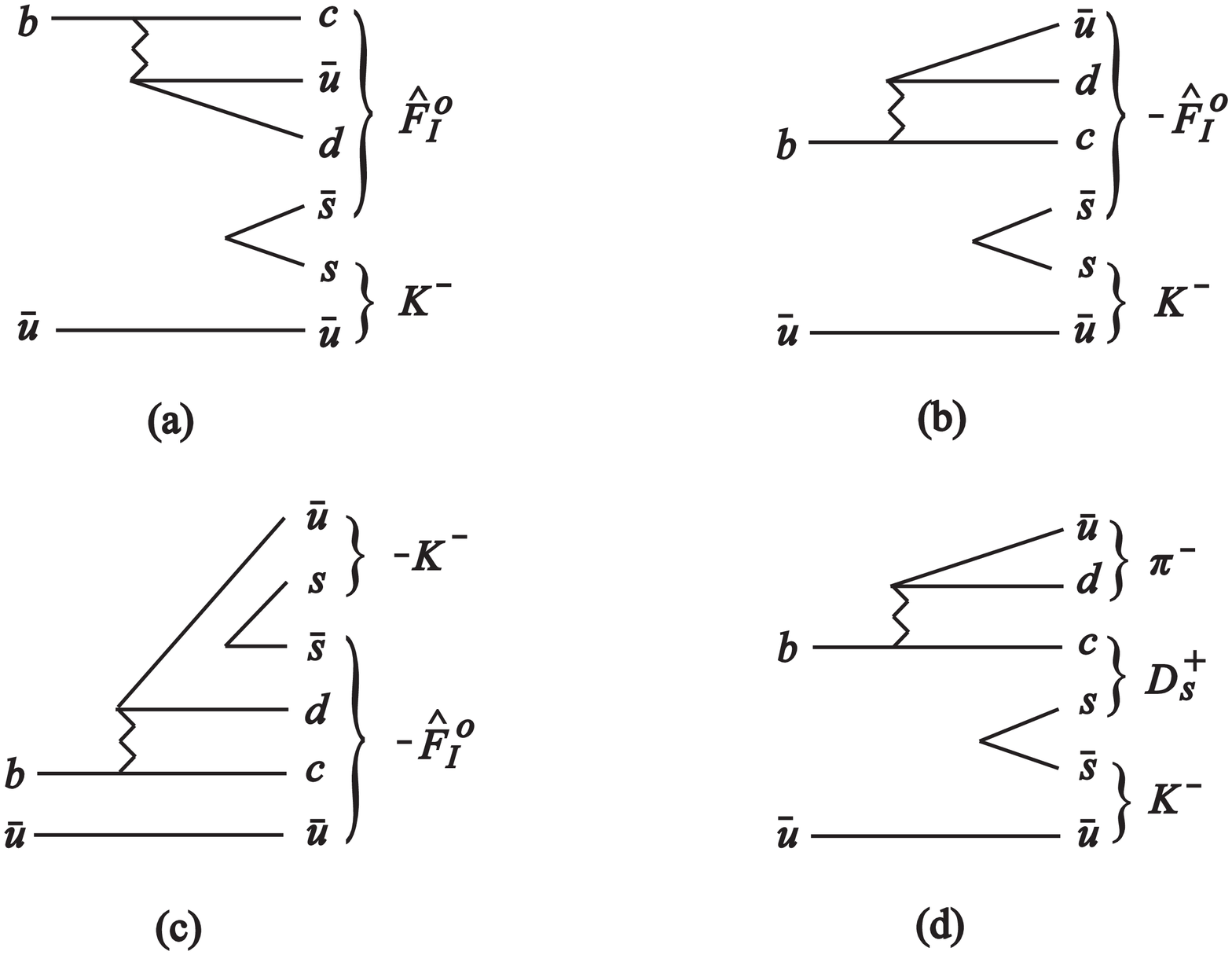}
\label{fig:barBu-8.eps}     %
%\end{center} 
%\caption{     
\begin{quote}  
Fig.~4. Production of charm-strange scalar mesons in weak decays of the 
$B_u^-$ meson. (a), (b) and (c) describe the production of $\hat F_I^0$ 
with $K^-$, and (d) the production of $D_s^+\pi^-$ with $K^-$.    
%}  
\end{quote}    
\end{figure}%        
%%%%%%%%%%%%%%%%%%%%%%%%%%%%%%%%%%%%%%%%%%%%%%%%%%%%%%%%%%%%%%%%%%%%%%%%
Although the diagrams Figs.~5(a) and (b) also yield the production of 
$\hat F_I^0$ with $\bar K^0$, they again interfere destructively, as in 
the above case.  

Because the process $B_u^+\rightarrow D^-\hat F_I^{++}$ is depicted by 
the same type of diagram as $B_u^+\rightarrow \bar D^0\hat F_I^+$, as 
seen above, the branching fraction for $\hat F_I^{++}$ production can be 
estimated as 
%%%%%%%%%%%%%%%%%%%%%%%%%%%%%%%%%%%%%%%%%%%%%%%%%%%%%%%%%%%%%%%%%%%%%%%%
\begin{eqnarray} 
&& \hspace{-5mm} 
{B}(B_u^+\rightarrow D^-\hat F_I^{++})   \nonumber\\
&&\sim {B}(B_u^+\rightarrow \bar D^0
\tilde D_{s0}^+(2317)[D_s^+\pi^0])_{\rm BABAR}
\nonumber\\
&&  \hspace{18mm} 
= (1.0\pm 0.3\pm 0.1^{+0.4}_{-0.2})\times 10^{-3}.  
                                              \label{eq:double-charge}
\end{eqnarray}
%%%%%%%%%%%%%%%%%%%%%%%%%%%%%%%%%%%%%%%%%%%%%%%%%%%%%%%%%%%%%%%%%%%%%%%%
In addition, the production of $\hat F_I^0$ is described by Fig.~3(a). 
This diagram is of the same type as that in Fig.~3(b), which depicts  
$B_d\rightarrow D^-\hat F_I^+$. Hence, the branching fraction for 
$\hat F_I^0$ production can be crudely estimated as 
%%%%%%%%%%%%%%%%%%%%%%%%%%%%%%%%%%%%%%%%%%%%%%%%%%%%%%%%%%%%%%%%%%%%%%%%
\begin{eqnarray} 
&& \hspace{-5mm} 
{B}(B_d^0\rightarrow \bar D^0\hat F_I^{0})   \nonumber\\
&& 
\sim {B}(B_d^0\rightarrow D^-
\tilde D_{s0}^+(2317)[D_s^+\pi^0])_{\rm BABAR}
\nonumber\\
&&  \hspace{18mm} 
= (1.8\pm 0.4\pm 0.3^{+0.6}_{-0.4})\times 10^{-3}. 
                                              \label{eq:neutral-1}
\end{eqnarray}
%%%%%%%%%%%%%%%%%%%%%%%%%%%%%%%%%%%%%%%%%%%%%%%%%%%%%%%%%%%%%%%%%%%%%%%%
In Eqs.~(\ref{eq:double-charge}) and (\ref{eq:neutral-1}), 
the last equalities were obtained in the BABAR 
experiment~\cite{BABAR-D_{s0}-B}. 

The BELLE collaboration~\cite{BELLE-D_{s0}-K^-} has observed the 
$\bar B_d^0\rightarrow K^-\tilde D_{s0}^+(2317)[D_s^+\pi^0]$ decay, as 
depicted in Fig.~5(c), and found 
%%%%%%%%%%%%%%%%%%%%%%%%%%%%%%%%%%%%%%%%%%%%%%%%%%%%%%%%%%%%%%%%%%%%%%%%
\begin{eqnarray}
%\begin{equation}
&&\hspace{-5mm}
{B}(\bar B_d^0\rightarrow K^-\tilde D_{s0}^+(2317))\cdot
{B}(\tilde D_{s0}^+(2317)\rightarrow D_s^+\pi^0) \nonumber\\
&&\hspace{20mm}
=(5.3^{+1.5}_{-1.3}\pm 0.7\pm 1.4)\times 10^{-5}.    
                                           \label{eq:BELLE-K}
%\end{equation}
\end{eqnarray}
%%%%%%%%%%%%%%%%%%%%%%%%%%%%%%%%%%%%%%%%%%%%%%%%%%%%%%%%%%%%%%%%%%%%%%%%%
%%%%%%%%%%%%%%%%%%%%%%%%%%%%%%%%%%%%%%%%%%%%%%%%%%%%%%%%%%%%%%%%%%%%%%%%
\begin{figure}[!t]    %
%\begin{center}    %\hspace{-20mm}
\includegraphics[width=70mm,clip]{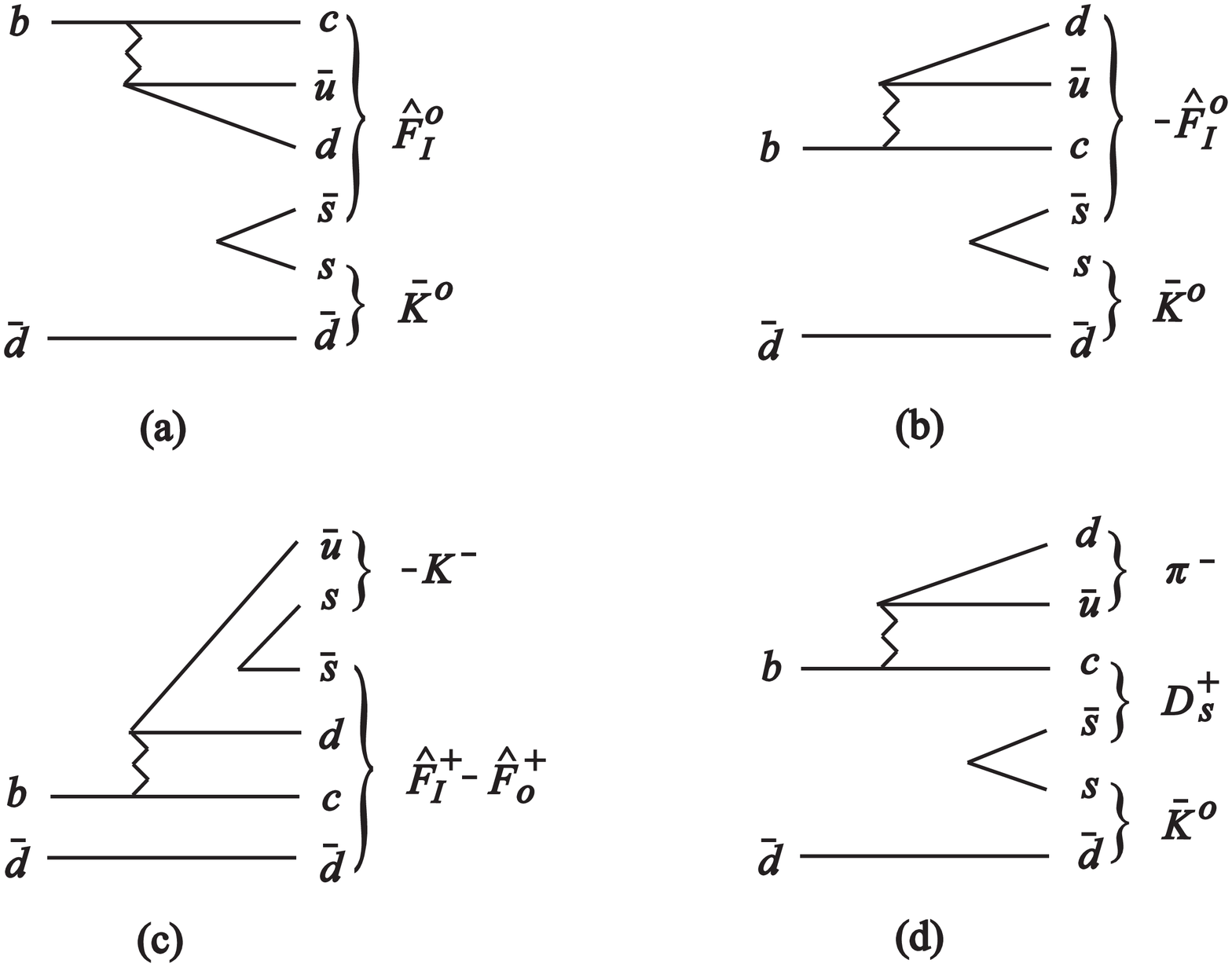}
\label{fig:barBd-8.eps}                %
%\end{center} 
%\caption{     
\begin{quote} 
Fig.~5. Production of charm-strange scalar mesons in weak decays of the 
$\bar B_d^0$ meson. (a) and (b) describe the production of $\hat F_I^0$ 
with $\bar K^0$, (c) the production of $\hat F_I^+$ and 
$\hat F_0^+$ with $K^-$, and (d) the production of $D_s^+\pi^-$ with 
$\bar K^0$.     
%}  
\end{quote}    
\end{figure}%        
%%%%%%%%%%%%%%%%%%%%%%%%%%%%%%%%%%%%%%%%%%%%%%%%%%%%%%%%%%%%%%%%%%%%%%%%
Identifying the above $\tilde D_{s0}^+(2317)$ with $\hat F_I^+$ and 
taking ${B}(\tilde D_{s0}^+(2317)\rightarrow D_s^+\pi^0)\simeq 100$ \%, 
as mentioned above,  we obtain the crude estimate 
%%%%%%%%%%%%%%%%%%%%%%%%%%%%%%%%%%%%%%%%%%%%%%%%%%%%%%%%%%%%%%%%%%%%%%%%%%
\begin{equation}
{B}(\bar B_d^0\rightarrow K^-\hat F_I^+)\sim 
10^{-4} - 10^{-5}. 
                                                     \label{eq:input}
\end{equation}
%%%%%%%%%%%%%%%%%%%%%%%%%%%%%%%%%%%%%%%%%%%%%%%%%%%%%%%%%%%%%%%%%%%%%%%%%
This is of the same order as (or slightly smaller than) the measured 
branching fraction~\cite{BELLE-D_{s0}-K^-} for a typical color mismatched decay,  
%%%%%%%%%%%%%%%%%%%%%%%%%%%%%%%%%%%%%%%%%%%%%%%%%%%%%%%%%%%%%%%%%%%%%%%%%
${B}(\bar B_d^0\rightarrow D^0\pi^0)_{\rm BELLE}=(2.31\pm 0.12\pm 0.23)
\times 10^{-4}$. 
%%%%%%%%%%%%%%%%%%%%%%%%%%%%%%%%%%%%%%%%%%%%%%%%%%%%%%%%%%%%%%%%%%%%%%%%%
This seems reasonable, because both of these decays involve 
rearrangements of colors before going to the final states, and because 
the former includes an $s\bar s$ pair creation. 
Using Eq.~(\ref{eq:input}) as the input data, we now estimate the 
branching fraction for production of $\hat F_I^0$. 
The $B_u^-\rightarrow K^-\hat F_I^0$ decay is depicted in Fig.~4(c), 
which is of the same type as Fig.~5(c) describing the decay 
$\bar B_d^0\rightarrow K^-\hat F_I^+$. As $\hat F_I^+$ is identified 
with $\tilde D_{s0}^+(2317)[D_s^+\pi^0]$, 
it is expected that 
%%%%%%%%%%%%%%%%%%%%%%%%%%%%%%%%%%%%%%%%%%%%%%%%%%%%%%%%%%%%%%%%%%%%%%%%%
\begin{eqnarray}
%\begin{equation}
&&\hspace{-10mm}
{B}(B_u^-\rightarrow K^-\hat F_I^0)   \nonumber\\
&&
\sim {B}(\bar B_d^0\rightarrow K^-\hat F_I^+) 
\sim 10^{-4} - 10^{-5}, 
\label{eq:neutral-2}
%\end{equation}
\end{eqnarray}
%%%%%%%%%%%%%%%%%%%%%%%%%%%%%%%%%%%%%%%%%%%%%%%%%%%%%%%%%%%%%%%%%%%%%%%%
if the contributions depicted in Figs.~4(a) and (b) cancel. 

Next, we consider the search for the iso-singlet $\hat F_0^+$ meson. 
Although $\hat F_I^+$ and $\hat F_0^+$ can be produced in $B$ decays 
described by the same diagrams, $\hat F_I^+$ decays dominantly into 
$D_s^+\pi^0$, but the radiative $D_s^{*+}\gamma$ decay is strongly 
suppressed, so that the assignment of $D_{s0}^+(2317)$ to $\hat F_I^+$ 
is consistent with Eq.~(\ref{eq:CLEO}). By contrast, in the case 
of $\hat F_0^+$, its radiative decay is much stronger than the isospin 
non-conserving $D_s^+\pi^0$ decay; i.e., we have 
%%%%%%%%%%%%%%%%%%%%%%%%%%%%%%%%%%%%%%%%%%%%%%%%%%%%%%%%%%%%%%%%%%%%%%%%%
${B}(\hat F_0^+\rightarrow D_s^{*+}\gamma) 
\gg {B}(\hat F_0^+\rightarrow D_s^{+}\pi^0)$. 
%%%%%%%%%%%%%%%%%%%%%%%%%%%%%%%%%%%%%%%%%%%%%%%%%%%%%%%%%%%%%%%%%%%%%%%%%
Therefore, if the masses of $\hat F_0^+$ and $\hat F_I^+$ are 
approximately equal and they are produced in $B$ decays represented by 
the same diagrams, Figs.~2(a), 3(b) and 5(c), it should be possible to 
observe them as resonances with nearly equal masses in the two channels 
$D_s^+\pi^0$ and $D_s^{*+}\gamma$. In fact, the BELLE 
experiment~\cite{BELLE-D_{s0}} has observed an indication of a 
resonance peak degenerate with $D_{s0}^+(2317)$ in the $D_s^{*+}\gamma$ 
channel, as well as the $D_s^+\pi^0$ channel. 

Although the CDF collaboration also has studied spectra of 
$D_s^+\pi^\pm$ produced inclusively from the Tevatron, neutral and 
doubly charged partners of $D_{s0}^+(2317)$ have not been observed. 
In this case, however, it is believed that very large numbers of 
background $D_s^+\pi^\pm$ events are produced, because the beam energy 
is very high. Therefore, it would be very difficult to extract the 
signal of $\hat F_I^0\rightarrow D_s^+\pi^\pm$ events in this kind of 
experiment. 

In summary, we have studied the production of charm-strange scalar 
four-quark mesons through hadronic weak decay. For this purpose, we have 
drawn quark-line diagrams within the minimal $q\bar q$ pair creation 
and have found that detecting neutral and doubly charged partners of 
$D_{s0}^+(2317)$ in inclusive $e^+e^-\rightarrow c\bar c$ is likely 
quite difficult, although $D_{s0}^+(2317)$ itself has already been 
observed. Taking these points into consideration, we have studied the 
possibility of their detection in hadronic weak decays of $B$ mesons. 
We have estimated the branching fractions for decays of $B$ mesons 
producing $\hat F_I^{++}$ and  $\hat F_I^0$ as 
%%%%%%%%%%%%%%%%%%%%%%%%%%%%%%%%%%%%%%%%%%%%%%%%%%%%%%%%%%%%%%%%%%%%%
${B}(B_u^+\rightarrow D^-\hat F_I^{++})
\sim {B}(B_d^0\rightarrow \bar D^0\hat F_I^{0}) \sim 10^{-3}$ and 
${B}(B_u^-\rightarrow K^-\hat F_I^{0})\sim  10^{-4} - 10^{-5}$. 
%%%%%%%%%%%%%%%%%%%%%%%%%%%%%%%%%%%%%%%%%%%%%%%%%%%%%%%%%%%%%%%%%%%%%%
Singly charged $\hat F_I^+$ and $\hat F_0^+$ are produced in hadronic 
weak decays of $B$ mesons described by the same diagrams. However, 
$\hat F_I^+$ decays dominantly into $D_s^+\pi^0$, while the 
$\hat F_0^+\rightarrow D_s^+\pi^0$ decay is much weaker than the 
$\hat F_0^+\rightarrow D_s^{*+}\gamma$. Therefore, we conclude that 
$\hat F_I^+$ and $\hat F_0^+$ could be observed as resonances with 
approximately equal masses in two different channels, $D_s^+\pi^0$ and 
$D_s^{*+}\gamma$, as the BELLE collaboration observed. 

%%%%%%%%%%%%%%%%%%%%%%%%%%%%%%%%%%%%%%%%%%%%%%%%%%%%%%%%%%%%%%%%%%%%%%%
\section*{Acknowledgements} 
The author would like to thank Prof. T.~Onogi and Prof. Y.~Kanada-En'yo 
of the Yukawa Institute for Theoretical Physics, Kyoto University, for 
valuable discussions and comments. He is also grateful to Prof.~K.~Abe, 
KEK, for informing him of the present status of experimental searches 
for tetra-quark mesons, and Prof.~H.~Terao and Prof.~T.~Izubuchi, 
Kanazawa University, for encouragement. This work is supported in part 
by a Grant-in-Aid for Science Research from the Ministry of Education, 
Culture, Sports, Science and Technology of Japan 
(No. 16540243).
%%%%%%%%%%%%%%%%%%%%%%%%%%%%%%%%%%%%%%%%%%%%%%%%%%%%%%%%%%%%%%%%%%%%%%%

%%%%%%%%%%%%%%%%%%%%%%%%%%%%%%%%%

%\end{references}
%%%%%%%%%%%%%%%%%%%%%%%
\end{document}